\documentclass[aps,prl,reprint,superscriptaddress,longbibliography,nofootinbib]{revtex4-2}

\usepackage{graphicx}
\usepackage{amsmath,amsfonts,amssymb}
\usepackage{physics}
\usepackage{xcolor}
\usepackage{hyperref}
\hypersetup{colorlinks=true,citecolor=blue,linkcolor=blue,urlcolor=blue}

\graphicspath{{figure/}}

\begin{document}

\title{Ultrasound Evidence for a Low-Temperature Anomaly Inside the Superconducting State of 4Hb-TaS$_2$}

\author{Yongwei Li}
\email{y.w.li@sjtu.edu.cn}
\affiliation{Tsung-Dao Lee Institute \& School of Physics and Astronomy, Shanghai Jiao Tong University, Shanghai 201210, China}

\author{Dmitri V. Efremov}
\affiliation{Institute for Solid State Research, Leibniz IFW Dresden, Helmholtzstra{\ss}e 20, 01069 Dresden, Germany}

\author{Paul Leask}
\affiliation{Department of Physics, KTH Royal Institute of Technology, Stockholm SE-10691, Sweden}

\author{Andreas Hauspurg}
\affiliation{Dresden High Magnetic Field Laboratory (HLD-EMFL) and W\"urzburg-Dresden Cluster of Excellence ctd.qmat, Helmholtz-Zentrum Dresden-Rossendorf, 01328 Dresden, Germany}

\author{{Irena Feldman}}
\affiliation{Israel Institute of Technology, Haifa 32000, Israel}

\author{Jochen Wosnitza}
\affiliation{Dresden High Magnetic Field Laboratory (HLD-EMFL) and W\"urzburg-Dresden Cluster of Excellence ctd.qmat, Helmholtz-Zentrum Dresden-Rossendorf, 01328 Dresden, Germany}
\affiliation{Institute for Solid State and Materials Physics, Technische Universit\"at Dresden, 01062 Dresden, Germany}

\author{Amit Kanigel}
\affiliation{Israel Institute of Technology, Haifa 32000, Israel}

\author{Sergei Zherlitsyn}
\affiliation{Dresden High Magnetic Field Laboratory (HLD-EMFL) and W\"urzburg-Dresden Cluster of Excellence ctd.qmat, Helmholtz-Zentrum Dresden-Rossendorf, 01328 Dresden, Germany}

\author{Hans-Henning Klauss}
\affiliation{Institute for Solid State and Materials Physics, Technische Universit\"at Dresden, 01062 Dresden, Germany}

\author{Vadim Grinenko}
\email{vadim.grinenko@sjtu.edu.cn}
\affiliation{Tsung-Dao Lee Institute \& School of Physics and Astronomy, Shanghai Jiao Tong University, Shanghai 201210, China}

\begin{abstract}
We report low-temperature ultrasound measurements on single crystals of the layered van der Waals superconductor 4Hb-TaS$_2$. Specific heat and ac magnetic susceptibility show a sharp bulk superconducting transition at \(T_{\rm c}\approx 2.9\)~K. Ultrasound measurements reveal an additional anomaly deep inside the superconducting state near \(T^{*}\approx 1\)~K. The most direct signature is observed in the relative ultrasonic attenuation change \(\Delta\alpha\): instead of being rapidly suppressed at \(T_{\rm c}\), \(\Delta\alpha\) remains large throughout the intermediate superconducting regime and drops strongly only near \(T^{*}\). This loss of acoustic dissipation is accompanied by a pronounced anomaly in the relative sound velocity change \(\Delta v/v\), indicating strong coupling to the lattice. The low-temperature anomaly is rapidly suppressed by magnetic field and by Se substitution, suggesting a possible superconducting origin of the anomaly. We speculate that this feature may be related to induced superconductivity in the 1T layers.
\end{abstract}

\maketitle

Layered transition-metal dichalcogenides provide a rich setting for studying superconductivity in the presence of charge order, lattice symmetry breaking, and electronic correlations~\cite{Manzeli2017}. 4Hb-TaS$_2$ is especially interesting because it forms a natural van der Waals heterostructure composed of alternating 1T-TaS$_2$ and 1H-TaS$_2$ layers~\cite{DISALVO1973,ribak2020chiral}. The 1T-derived layers are associated with star-of-David charge-density-wave order and strong electronic correlations~\cite{Law2017,Ribak2017,Murayama2020}, whereas superconductivity is primarily associated with the metallic 1H-derived layers. This alternating structure provides a natural platform in which superconductivity may couple to additional low-energy degrees of freedom.

Several experiments have suggested that superconductivity in 4Hb-TaS$_2$ is unconventional. Muon spin rotation measurements reported enhanced relaxation in the superconducting state~\cite{ribak2020chiral}. Little-Parks measurements observed phase shifts suggestive of an unconventional superconducting response~\cite{Almoalem2024}.

Measurements of the in-plane upper critical field revealed a pronounced twofold anisotropy near \(T_{\rm c}\), and scanning tunneling microscopy (STM) found stripe-like conductance modulations that lower the apparent rotational symmetry~\cite{Silber2024}. On the other hand, at low temperature STM finds round vortex cores suggesting an isotropic order parameter \cite{Nayak2021a}. These observations motivate a basic experimental question: does the superconducting state formed at \(T_{\rm c}\) evolve smoothly down to the lowest temperatures, or does it show an additional low-temperature anomaly?

Ultrasound is well-suited to address this question because it probes both the elastic response and acoustic dissipation. The sound velocity is sensitive to changes in elastic moduli and to the coupling between strain and order parameters, while the attenuation is sensitive to low-energy excitations that absorb acoustic energy~\cite{Batlogg1985,benhabib2021ultrasound,Ghosh2021,Halcrow2024,Kurihara2017}. In a conventional fully gapped superconductor, the electronic contribution to ultrasonic attenuation is expected to decrease below \(T_{\rm c}\). Therefore, a large residual attenuation inside the superconducting state provides evidence for an additional dissipative channel that is not fully removed by the main superconducting transition.

Here, we present ultrasound measurements revealing an anomaly within the superconducting state of 4Hb-TaS$_2$. The main superconducting transition occurs at \(T_{\rm c}\approx 2.9\)~K and is clearly detected by specific heat, ac magnetic susceptibility, and sound velocity. A second anomaly appears near \(T^{*}\approx 1\)~K and is clearly resolved in both the ultrasound velocity and the relative attenuation \(\Delta\alpha\). 

Single crystals of 4Hb-TaS$_2$ were grown by the chemical vapor transport method as described previously~\cite{ribak2020chiral}. For ultrasound measurements, we used Se-substituted crystals with nominal compositions \(4\text{Hb-TaS}_{1.99}\text{Se}_{0.01}\) and \(4\text{Hb-TaS}_{1.98}\text{Se}_{0.02}\), referred to below as 1\% and 2\% Se-substituted samples. A small amount of Se substitution was used to stabilize crystals of sufficient size for ultrasound measurements while keeping the main superconducting transition close to 3~K.

We employed a pulse-echo technique to measure the relative sound velocity change \(\Delta v/v\) and the relative ultrasonic attenuation change \(\Delta\alpha\) in longitudinal and transverse acoustic modes. The sound propagated along the hexagonal \(b\) axis, corresponding to the [100] direction in the notation used here. The longitudinal mode probes primarily the \(C_{11}\) elastic response, while the transverse geometry probes the \(C_{66}\) shear response. The attenuation reported here is a relative quantity, not an absolute attenuation coefficient. For each measurement configuration \(i\), we define \(\Delta\alpha_i=-20\log[A_i/A_{4\,{\rm K}}]/l\), where \(A_i\) is the measured echo amplitude, \(A_{4\,{\rm K}}\) is the reference amplitude at 4~K, and \(l\) is the acoustic path length. This convention allows us to track changes in acoustic dissipation with temperature and field, while avoiding an overinterpretation of \(\Delta\alpha\) as an absolute attenuation coefficient.

The absolute sound velocities and elastic moduli were estimated from the sound travel time through the crystal using \(C=\rho v^2\). For the 2\% Se-substituted crystal, we obtained \(v_L \approx 4490 \pm 30\)~m/s and \(v_T \approx 2380 \pm 110\)~m/s, corresponding to \(C_{11} \approx 138 \pm 2\)~GPa and \(C_{66} \approx 39 \pm 4\)~GPa, respectively.

\begin{figure}[t]
    \centering
    \includegraphics[width=\columnwidth]{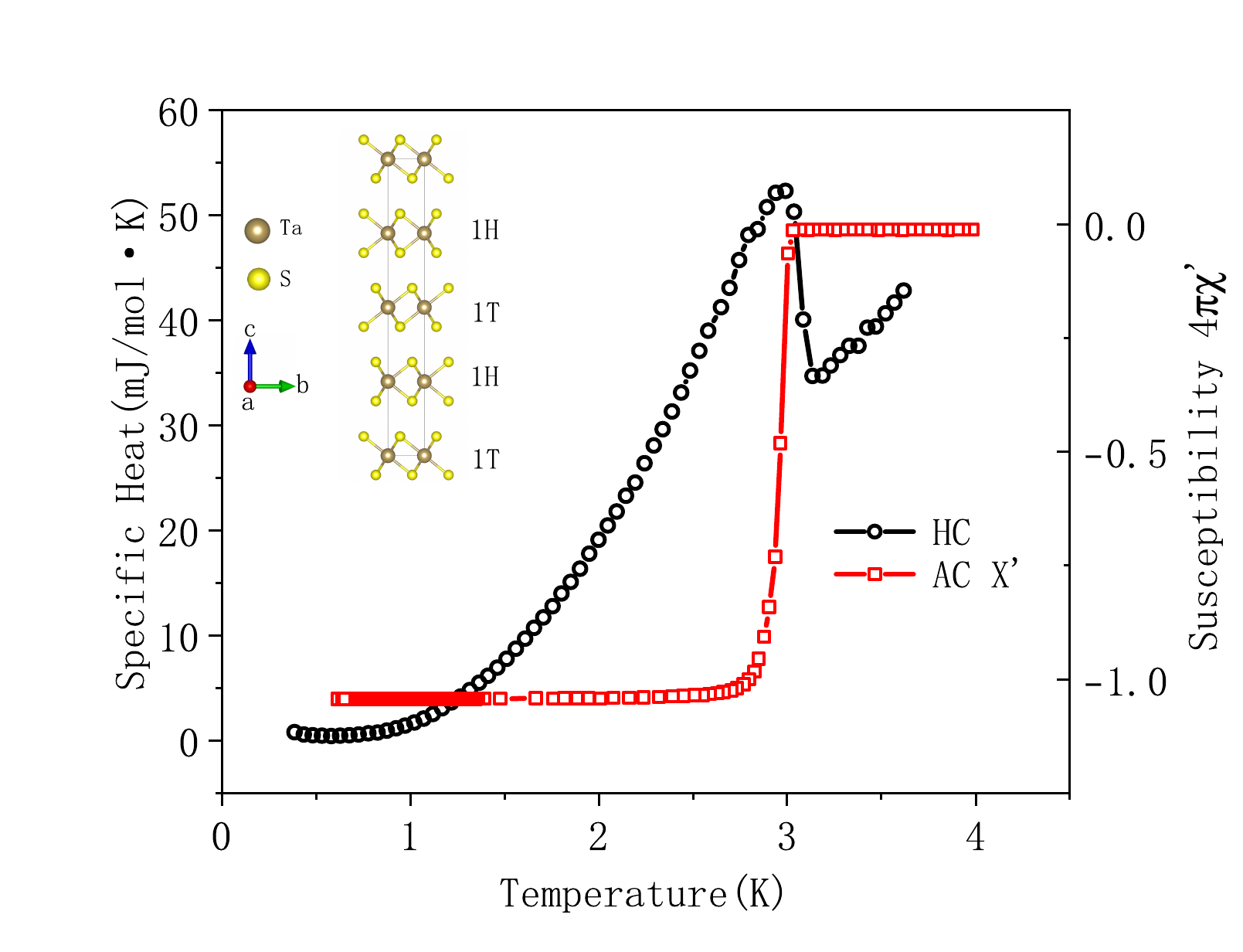}
    \caption{
    Sample characterization of 4Hb-TaS$_2$.
    Temperature dependence of the specific heat and the real part of the ac magnetic susceptibility for a \(4\text{Hb-TaS}_{1.99}\text{Se}_{0.01}\) single crystal measured in zero static magnetic field. Both probes show a sharp bulk superconducting transition at \(T_{\rm c}\approx 2.9\)~K. No comparably sharp second anomaly is resolved in these thermodynamic and magnetic probes down to the lowest measured temperatures. The inset shows the crystal structure of 4Hb-TaS$_2$, consisting of alternating 1T and 1H layers.
    }
    \label{fig:sample}
\end{figure}

Figure~\ref{fig:sample} summarizes the sample characterization. The specific heat and ac magnetic susceptibility both show a sharp superconducting transition near \(T_{\rm c}\approx 2.9\)~K, confirming bulk superconductivity and good sample quality. Within the resolution of these measurements, no comparably sharp second anomaly is resolved near 1~K. This observation is important for interpreting the ultrasound data below. The low-temperature anomaly is not a large entropy anomaly like the main superconducting transition. Instead, it is most strongly expressed in the elastic and dissipative response, indicating strong coupling to strain and to low-energy acoustic absorption.

\begin{figure*}[t]
  \centering
  \includegraphics[width=0.7\textwidth]{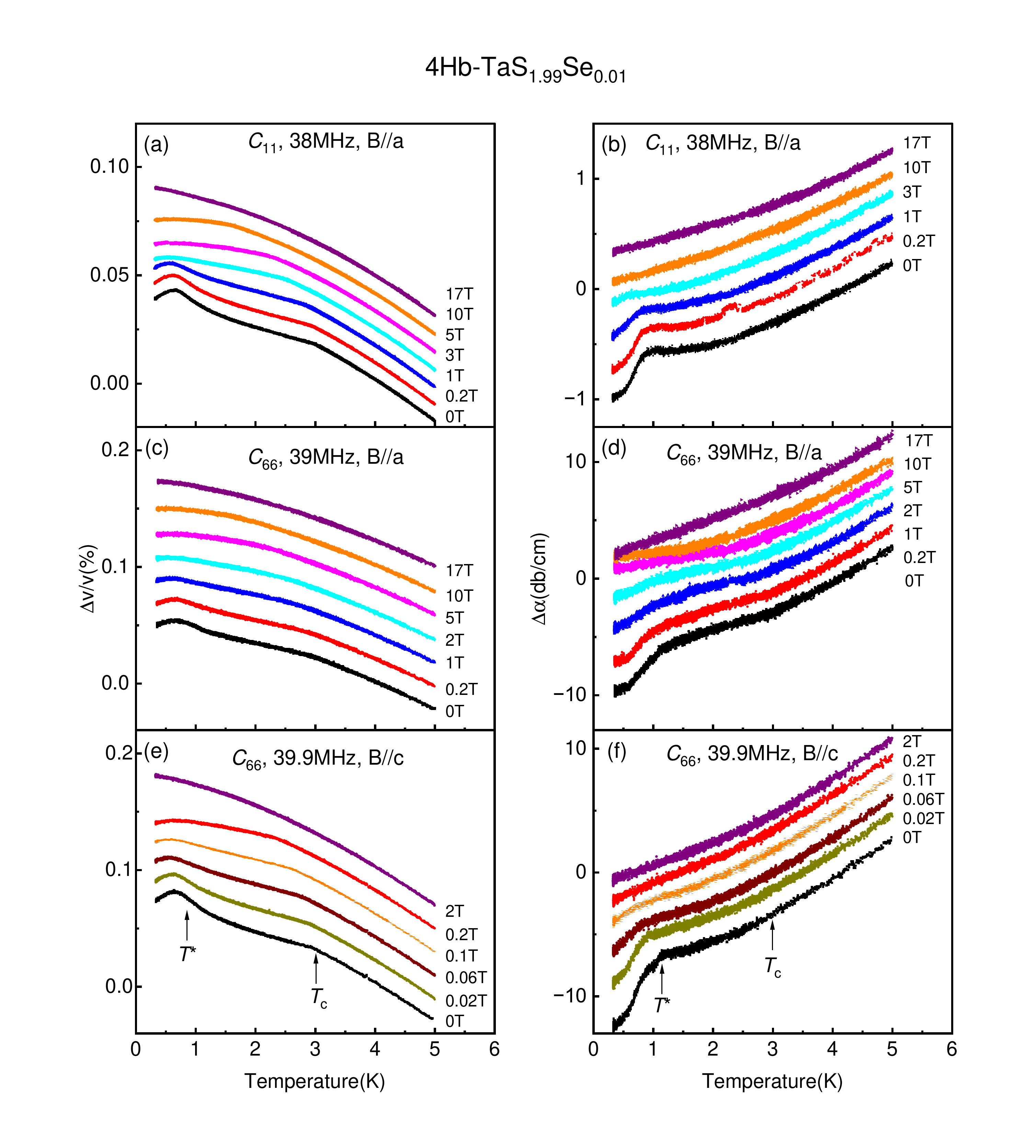}
  \caption{
  Ultrasound response of the 1\% Se-substituted \(4\text{Hb-TaS}_{1.99}\text{Se}_{0.01}\) sample.
  \textbf{(a,b)} Longitudinal \(C_{11}\) mode at \(f=38\)~MHz with magnetic field applied along the hexagonal \(a\) axis, \(B\parallel a\).
  \textbf{(c,d)} Transverse \(C_{66}\) mode at \(f=39\)~MHz with \(B\parallel a\).
  \textbf{(e,f)} Transverse \(C_{66}\) mode at \(f=39.9\)~MHz with magnetic field applied parallel to the \(c\) axis, \(B\parallel c\).
  Panels \textbf{(a,c,e)} show the relative sound velocity change \(\Delta v/v\), while panels \textbf{(b,d,f)} show the relative ultrasonic attenuation change \(\Delta\alpha\). Two features are resolved: the main superconducting transition at \(T_{\rm c}\approx 2.9\)~K and a second low-temperature anomaly near \(T^{*}\approx 1\)~K. The region between \(T_{\rm c}\) and \(T^{*}\) is referred to as the intermediate superconducting regime, while the region below \(T^{*}\) is referred to as the low-temperature superconducting regime. The curves are deliberately shifted along the $y$-axis for clarity.
  }
  \label{fig:se1}
\end{figure*}

We now turn to the ultrasound response of the 1\% Se-substituted sample, where the low-temperature anomaly is most pronounced. Figure~\ref{fig:se1} shows \(\Delta v/v\) and \(\Delta\alpha\) for longitudinal and transverse modes under magnetic field. The first clear feature is the main superconducting transition near \(T_{\rm c}\approx 2.9\)~K. It appears as a kink-like anomaly in \(\Delta v/v\) and shifts to lower temperature under magnetic field, as expected for a superconducting transition.

A second feature appears well inside the superconducting state near \(T^{*}\approx 1\)~K. This anomaly is present in the sound velocity and is particularly clear in the attenuation. In the transverse \(C_{66}\) response, \(\Delta\alpha\) remains large below \(T_{\rm c}\) and then drops sharply only near \(T^{*}\). This behavior is qualitatively different from the standard expectation for a simple fully gapped superconducting transition, where the attenuation should be reduced already below \(T_{\rm c}\). The observation, therefore, shows that the superconducting regime just below \(T_{\rm c}\) doesn't affect the low-energy excitations absorbing acoustic energy and the lower-temperature anomaly at \(T^{*}\) is associated with a sharp loss of this dissipation. 

\begin{figure*}[t]
  \centering
  \includegraphics[width=0.7\textwidth]{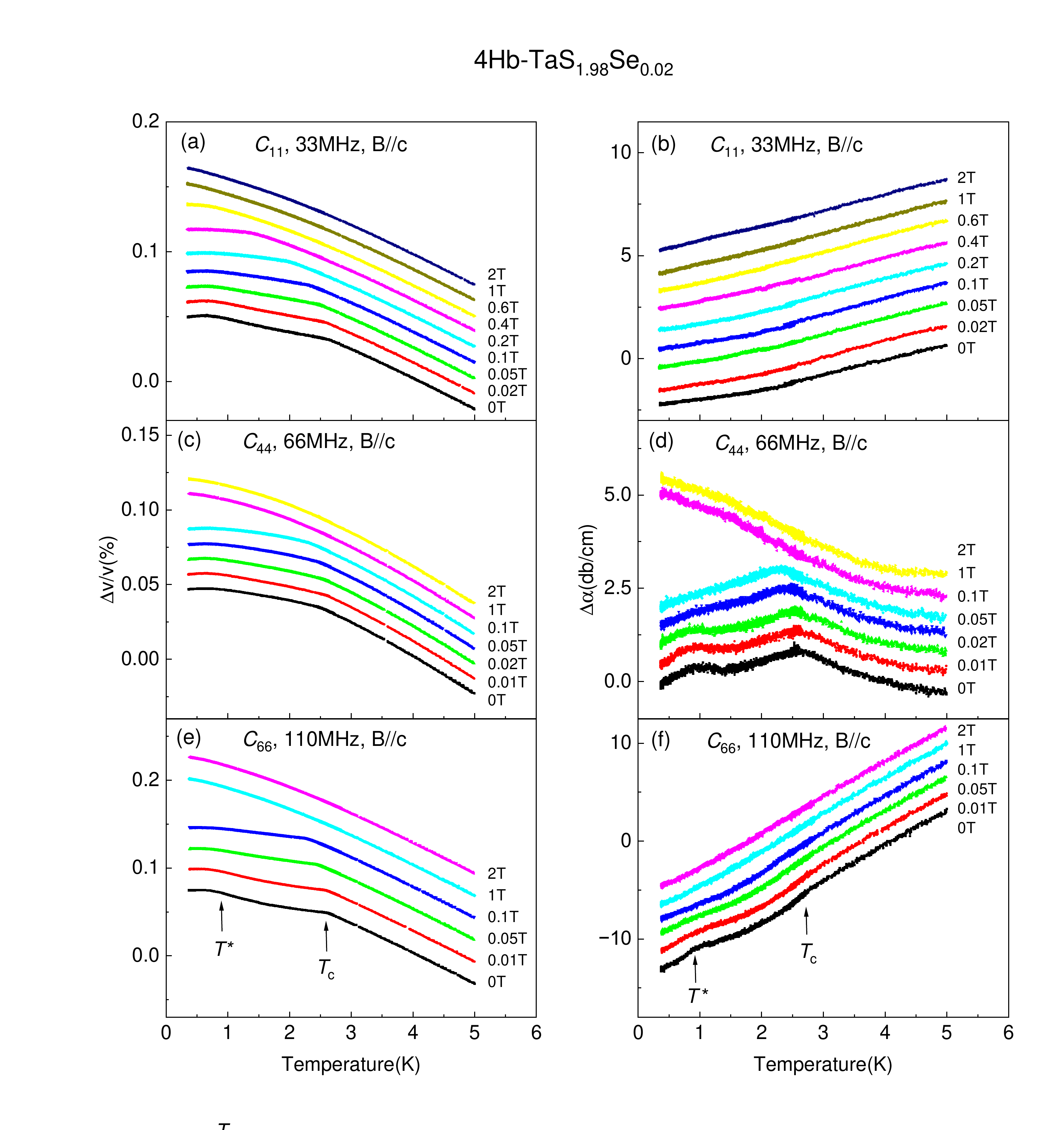}
  \caption{
  Ultrasound response of the 2\% Se-substituted \(4\text{Hb-TaS}_{1.98}\text{Se}_{0.02}\) sample with magnetic field applied parallel to the \(c\) axis, \(B\parallel c\).
  \textbf{(a,c,e)} Temperature dependence of the relative sound velocity change \(\Delta v/v\) for the \(C_{11}\), \(C_{44}\), and \(C_{66}\) acoustic modes, respectively.
  \textbf{(b,d,f)} Corresponding relative ultrasonic attenuation change \(\Delta\alpha\) for the same modes.
  Compared with the 1\% Se-substituted sample, the low-temperature anomaly near \(T^{*}\approx 1\)~K is strongly weakened and broadened, while the main superconducting transition remains clearly visible. The curves are deliberately shifted along the $y$-axis for clarity.
  }
  \label{fig:se2}
\end{figure*}

Figure~\ref{fig:se2} shows the corresponding data for the 2\% Se-substituted sample. The main superconducting transition is well resolved in both the ultrasound velocity and attenuation, although \(T_{\rm c}\) is slightly reduced. By contrast, the low-temperature anomaly near \(T^{*}\) is strongly weakened. These data show that Se doping suppresses both the low-energy excitations responsible for the ultrasonic dumping above $T^*$, and the low-temperature anomaly at $T^*$, suggesting a possible relation between them. 

The experimental findings can be summarized in terms of two distinct acoustic responses. In the intermediate superconducting regime, the system is superconducting but remains acoustically dissipative. This means that the superconducting gap associated with the main transition does not eliminate the low-energy modes that couple to sound. Below \(T^{*}\), the attenuation drops sharply, and the sound velocity shows a strong anomaly. The anomaly near \(T^{*}\) therefore corresponds to a loss or reorganization of the dissipative spectral weight. From the ultrasound perspective, the essential difference between the intermediate and low temperature regime is the presence or absence of an additional strain-coupled dissipative channel. 

\begin{figure}[t]
    \centering
    \includegraphics[width=\columnwidth]{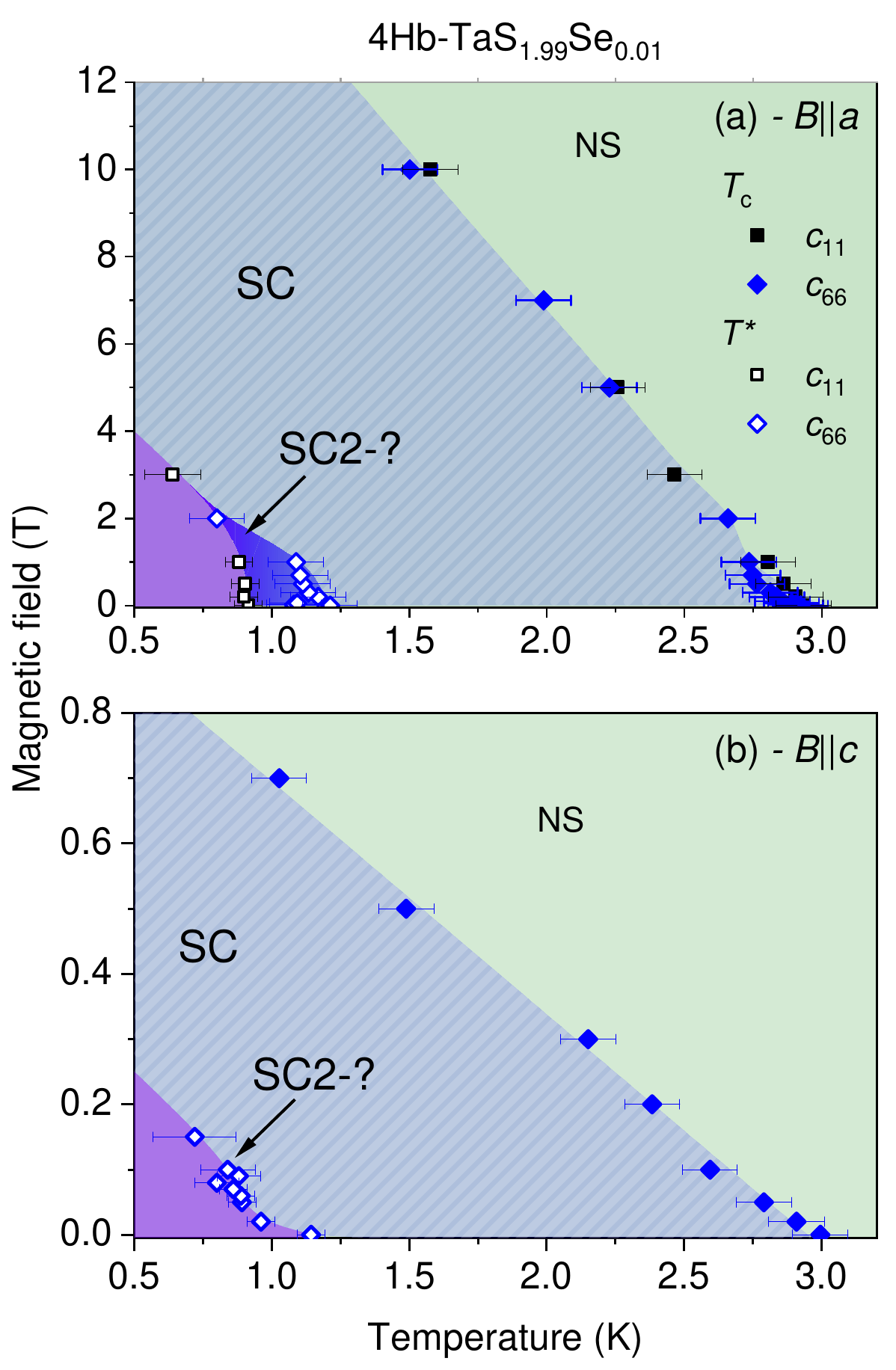}
    \caption{
    Temperature--magnetic-field summary of characteristic temperatures derived from ultrasound measurements on the 1\% Se-substituted sample.
    \textbf{(a)} Magnetic field applied along the hexagonal \(a\) axis, \(B\parallel a\).
    \textbf{(b)} Magnetic field applied parallel to the \(c\) axis, \(B\parallel c\).
    The upper line corresponds to the main superconducting transition at \(T_{\rm c}\). The lower line tracks the low-temperature ultrasound anomaly at \(T^{*}\). Symbols denote characteristic temperatures extracted from anomalies in different acoustic modes.
    }
    \label{fig:phase}
\end{figure}

The field-temperature phase diagram for 1\% sample is shown in Fig.~\ref{fig:phase}. The upper line corresponds to the main superconducting transition at \(T_{\rm c}\) and follows the expected behavior of the superconducting upper critical field. The lower characteristic line tracks the low-temperature anomaly at \(T^{*}\). This lower feature is rapidly suppressed by the magnetic field for both field orientations and disappears at fields well below the upper critical field.
The strong-field sensitivity points to a possible superconducting or magnetic nature of the second transition. However, the latter seems unlikely, as no indication of magnetism was found in the magnetic susceptibility measurements (Fig.~\ref{fig:sample}).  

The ultrasound data do not uniquely identify the microscopic origin associated with the lower anomaly, but they impose several constraints. The mechanism must allow superconductivity to appear at \(T_{\rm c}\) while leaving a strong acoustic dissipation channel active in the intermediate superconducting regime. It must also produce a sharp loss of dissipation near \(T^{*}\), couple strongly to lattice strain, and be sensitive to weak disorder and magnetic field. In addition, the transition at $T^*$ is invisible in the specific heat and magnetic susceptibility (Fig.~\ref{fig:sample}), suggesting a small associated entropy.

One possible explanation is multicomponent superconductivity. In this scenario, the transition at \(T_{\rm c}\) establishes the first superconducting component, while an additional component or a change in the superconducting order-parameter structure develops 
at a lower temperature. Such a change could reduce the low-energy quasiparticle spectral weight and produce an additional elastic anomaly through coupling to strain. However, a missing anomaly in the specific heat at $T^{*}$
and, in contrast, a much stronger ultrasound response at $T^{*}$ than at the primary superconducting transition \(T_{\rm c}\) makes this scenario unlikely. 

A second possibility is band- or layer-selective superconductivity, in which different electronic sectors acquire gaps or phase coherence at different characteristic temperatures. In such a case, \(T_{\rm c}\) would mark the onset of the dominant superconducting response in 1H layers, while \(T^{*}\) would correspond to an additional low-temperature reorganisation of a weaker sector, for example, induced superconductivity in 1T layers, which can be metallic due to a charge transfer between 1T ans 1H layers suggested by recent ARPES data~\cite{Almoalem2024b}.

In summary, low-temperature ultrasound measurements reveal an additional anomaly inside the superconducting state of 4Hb-TaS$_2$. The main superconducting transition at \(T_{\rm c}\approx 2.9\)~K is confirmed by specific heat, ac magnetic susceptibility, and sound velocity. A second anomaly appears near \(T^{*}\approx 1\)~K inside the superconducting state. This lower anomaly is marked by a sharp reduction of the relative ultrasonic attenuation change \(\Delta\alpha\) and by a pronounced anomaly in the relative sound velocity change \(\Delta v/v\). It is rapidly weakened by magnetic field and by weak Se substitution. These results show that the superconducting state between \(T_{\rm c}\) and \(T^{*}\) contains an additional strain-coupled dissipative channel, while the lower-temperature regime below \(T^{*}\) is characterized by much weaker acoustic dissipation. The microscopic origin of this reorganization remains open, but the data establish that superconductivity in 4Hb-TaS$_2$ does not evolve as a single acoustically homogeneous state from \(T_{\rm c}\) to the lowest temperatures.

\begin{acknowledgments}
We acknowledge useful discussions with colleagues and support from the respective funding agencies and institutions. This work was supported by the German--Israeli Project Cooperation (DIP), Grant Nos. 529677299 and 3970/1-1. We acknowledge support from the Deutsche Forschungsgemeinschaft (DFG) through the W\"{u}rzburg-Dresden Cluster of Excellence on Complexity, Topology and Dynamics in Quantum Matter--$ctd.qmat$ (EXC 2147, Project No.\ 390858490), as well as support from the HLD at HZDR, which is a member of the European Magnetic Field Laboratory (EMFL). P.L. acknowledges the Olle Engkvists Stiftelse through Grant No. 226-0103.
\end{acknowledgments}

\bibliography{references}

\end{document}